# Nonlinear and Ultrafast All-Dielectric Metasurfaces at the Center for Integrated Nanotechnologies

Sylvain D. Gennaro[1,2], Raktim Sarma[1,2] and Igal Brener[1,2]

1. Center for Integrated Nanotechnologies, Sandia National Laboratories, Albuquerque NM, 87123, USA
2. Sandia National Laboratories, Albuquerque, NM, 87123, USA.

Abstract :

**Metasurfaces control optical wavefronts via arrays of nanoscale resonators laid out across a surface. When combined with III-V semiconductors with strong optical nonlinearities, a variety of nonlinear effects such as harmonic generation and all optical modulation can be enabled and enhanced at the nanoscale. This review presents our research on engineering and boosting nonlinear effects in ultrafast and nonlinear semiconductor metasurfaces fabricated at the Center for Integrated Nanotechnologies (CINT). We cover our recent works on parametric generation of harmonic light via direct and cascaded processes in GaAs-metasurfaces using Mie-like optical resonances or symmetric-protected Bound State in the Continuum, and then describe the recent advances on harmonic generation in all-dielectric metasurfaces coupled to intersubband transitions in III-V semiconductor heterostructures. The review concludes on the potential of metasurfaces to serve as the next platform for on-chip quantum light generation.**

## Introduction

In 1960, the world bore witness to the first experimental demonstration of lasing in a Ruby laser by Theodore H. Maiman. [1] This fascinating discovery provided the scientific community with an operating device capable of generating intense and coherent light beams, enabling the exploration of extreme light-matter interaction such as nonlinear frequency-mixing. [2] Soon enough, the generation of second harmonic in crystalline quartz was observed, [3] propelling the field of optics and photonics to a new era. Nowadays, nonlinear optical effects are ubiquitous in all aspects of modern science such as in bioimaging and chemical sensing, nanomedicine, material characterization, optical information processing, quantum optics, supercontinuum light generation and in the design of novel laser sources with optical parametric amplifiers.

For a long time, birefringent crystals and quasi-phase matched crystals and waveguides were the dominant media for nonlinear optical systems. [4] In recent years however, there has been a growing need for compact integrated photonics and optoelectronics to increase the system's performance and reduce its power consumption. For this purpose, conventional bulk optical elements are too large to meet these requirements both for linear and nonlinear optical applications. For the latter, the dispersive nature of traditional materials in conjunction with their inherently-weak nonlinear optical susceptibilities require a coherent accumulation of the nonlinear signal over the length of the system while compensating for any momentum mismatch between incident and converted light. [2] This limitation of bulk optical media has driven researchers to novel ways of miniaturizing optical components to minimal dimensions – reaching down to the nanoscale. [5–13] However, due to the smaller volume of nonlinear materials involved in nanophononics, it is challenging to design nonlinear optical systems with useful conversion efficiencies as phase-matching techniques of bulk optics are no longer applicable.

Over the past few years, nanoscale resonators arrayed as metasurfaces have represented a great advance in tackling this challenge. Due to their abilities to enhance electromagnetic fields locally via light confinement into the subwavelength dimensions of each resonator (or optical antennas), nonlinear conversion efficiencies have surpassed that of conventional crystals for an equivalent volume of nonlinear materials. [7,11,14–18] Historically, the optical antennas that constitute a metasurface were first fabricated with noble metals such as gold, silver or even aluminum [9,19,20] as they can support plasmon resonances at the surface of the metal (referred as localized surface plasmon resonances [21]). Tight electromagnetic field confinement at the metal surface enables the observation of second, [6,7,29,30,14,22–28] and third harmonic generation [28,31,32] in individual optical antennas. However, high ohmic losses, the associated thermal heating and their subsequent low damage threshold have limited their usefulness for nonlinear optics using high power lasers. Since the extrinsic efficiencies of nonlinear systems scales with some power-law of the incident power, it is paramount to tackle thermal losses in metasurfaces while simultaneously increasing the external efficiency.

In recent years, all-dielectric nanoparticles and metasurfaces made of semiconductors such as GaAs, [33–41] AlGaAs, [42–48] Si, [49–51] Ge, [52–54] GaP, [55] LiNbO3, [56] BatiO3 [57] or semiconductor heterostructures that support intersubband transition (IST) in multilayers quantum wells [58–67] have gained momentum in the field of nonlinear optics due to their low losses, and high field confinement within the volume of each nanoscale resonators. Each optical resonators can support localized Mie-type resonant modes (mostly electric and magnetic dipole modes [42,44,68–72] or anapole modes [73,74]) or high quality-factor optical modes such as bound state in the continuum (BIC). [34,46,75–82] Combined with resonant optical nonlinearities that can be supported by ISTs in semiconductor quantum wells [58–67], these metasurfaces can reach nonlinear conversion efficiencies on the order of a fraction of a percent for thicknesses less than 1/5[th] of the pump wavelength in vacuum. (see [67] for the highest efficiency reported to date), suitable to observe harmonic generation with much lower pump power without the need of high-intensity laser pulses.

Metasurfaces provide an unprecedent control over the light amplitude, phase and polarization of the optical fields inside the meta-atoms. This is due to the fact that most of the light conversion occurs within a subwavelength thickness of the constituent material of the metasurface. Phase-matching conditions are relaxed and all allowed nonlinear processes can, in principle, occur on an equal footing. For example, semiconductor metasurfaces have enabled the design of a dielectric metamixer that creates ten new frequencies across the visible and near-infrared spectrum, [35] and the observation of cascaded second order optical nonlinearities. [40]

In this review, we put in perspective the research done on ultrafast and nonlinear all-dielectric metasurfaces performed at the Center of Integrated Nanotechnologies (CINT) with respect to the existing literature in the field. We will first introduce the fundamental description of nonlinear effects at the nanoscale and introduce a general formalism on how to boost frequency mixing. Then, we will cover parametric generation of harmonic light in the GaAs-material platform developed at CINT. In a final section, we will present recent advances on non-parametric harmonic generation in metasurfaces coupled to intersubband transitions in n-doped $In_{0.53}Ga_{0.47}As/Al_{0.52}In_{0.48}As$ multi-QWs III-V semiconductor heterostructures.

## Theory of harmonic generation

Upon excitation with an electromagnetic wave, the dipole moments of electrons in natural materials polarize with respect to the orientation of the electric field, oscillating coherently, and leading to the emission of a scattered wave. [2] The motion of electrons in this case can be described by an Abraham-Lorentz harmonic oscillator model. [83] For large incident electromagnetic radiation, however, the electron trajectories start to deviate from that of a harmonic oscillator. In this regime, the optical response of a material becomes intensity-dependent and is no longer linear such that, in the case of frequency wave mixing – where photons of one frequency are converted to photons from another frequencies – multiple harmonics of the incident pump frequency can be generated. For example, as illustrated in figure 1a, third harmonic generation converts three photons of lower energies to one photon of higher energy. When multiple beams are involved, the output frequency is linked by an algebraic sum derived from energy conservation; for example, the output frequency, $\omega_2 = 2\omega - \omega_1$ of four wave mixing is given by the difference between two photons of frequency $\omega$, and one photon of frequency $\omega_1$. If the quantum state of the material is unchanged during the light conversion process, for example, there is no light absorption as in transparent lossless crystals, the process is said to be parametric (and instantaneous), and results from virtual energy states transitions. However, in recent years, scientists have designed metasurfaces that integrate engineered materials in which the nonlinear conversion process relies on resonant real energy states such as intersubband electronic transitions in InGaAs multi-quantum wells with AlInAs barriers. In this case, the material absorbs and reemit photons via interacting with real transitions, making the process non-instantaneous. This approach enables the access of much stronger nonlinear susceptibilities and does not need expensive high power and/or pulsed laser systems to achieve exceptional conversion efficiencies.

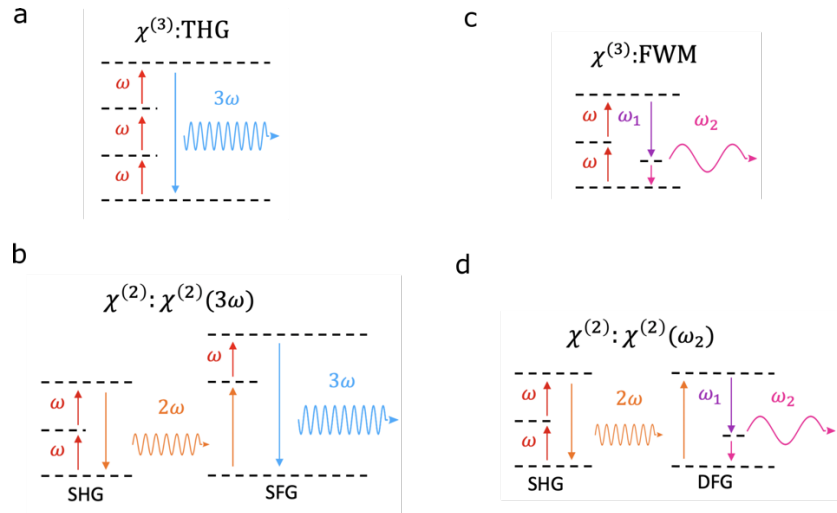

**Figure 1. Nonlinear frequency mixing processes.** Schematic of harmonic generation of a wave at frequencies $3\omega$ via (a) direct third harmonic generation $\chi^{(3)}$: $[\chi^{(3)}(-3\omega;\omega,\omega,\omega)]$ or (b) through cascaded second harmonic generation (SHG) and sum frequency generation (SFG). $[\chi^{(2)}(-2\omega;\omega,\omega):\chi^{(2)}(-3\omega;2\omega,\omega)]$; and at frequencies $\omega_2$ via (c) direct $\chi^{(3)}$ −FWM $[\chi^{(3)}(-\omega_2;\omega_1,\omega,\omega)]$ or (d) through cascaded SHG and DFG $[\chi^{(2)}(-2\omega;\omega,\omega):\chi^{(2)}(-\omega_2;\omega_1,2\omega)]$. Figures are adapted with permission from ref [40]. American Chemistry Society.

In principle, due to the absence of phase-matching, both direct (single step) and cascaded (consisting of several frequency mixing steps) nonlinear generation processes could exist, for example, as shown in figure 1a and c. Three wave mixing can arise from a direct $\chi^{(3)}$ − process or from a cascaded (sequential) process of $\chi^{(2)} - \chi^{(2)}$ of second harmonic generation and sum frequency generation. Four-wave mixing, one of the hallmarks of third order nonlinearities, can also be achieved by SHG of one pump beam, followed by difference frequency generation (DFG) with a second pump beam as shown in figure 1b and d. [4] At the nanoscale, cascaded optical nonlinearities have been observed in plasmonic optical antennas, [16] and in semiconductor metasurfaces. [40]

Each of these frequencies mixing processes is described by a nonlinear polarizability [2] such that, for second harmonic generation, we have:

$$P_2 \propto \iiint \chi^{(2)}(-2\omega;\omega,\omega)E(\omega)E(\omega) \tag{1}$$

Where $E(\omega)$ are the respective electromagnetic fields at the fundamental and harmonic frequencies and $\chi^{(2)}(-2\omega;\omega,\omega)$ is a third rank tensor, whose shape is subject to symmetry consideration of the material. For example, GaAs belongs to the point-group symmetry $\bar{4}3m$ such that only six non-zero elements of the tensor obey $\chi^{(2)}_{xyz} = \chi^{(2)}_{xzy} = \chi^{(2)}_{yxz} = \chi^{(2)}_{yzx} = \chi^{(2)}_{zxy} = \chi^{(2)}_{zyx}$. Experimental values of $\chi^{(2)}(-2\omega;\omega,\omega)$ for GaAs at visible wavelengths can be found in [84]

From equation 1, we can see that, to boost the nonlinear efficiencies of harmonic generation from metasurfaces, we need to find ways of enhancing electromagnetic fields at the nanoscale at the fundamental and harmonic frequencies. When working with narrow optical resonances though, it is important to consider the spectral bandwidth difference that might exist between the incident beam and the optical resonance it couples to as some of the pump energy might be reflected otherwise. [11,34] Another challenging but rewarding approach is to fabricate metasurfaces from materials with engineered resonant optical nonlinearities such as intersubband transitions (IST) in multilayers quantum wells. [58–67] This approach however also has its own limitations such as absorption at the fundamental and harmonic frequencies, which reduces the field enhancements at the pump and harmonic frequencies and thereby lowers the system's overall nonlinear conversion efficiency. We will cover these two approaches in the following sections.

# Frequency Conversion via Parametric Processes in All-Dielectric Metasurfaces

In metasurfaces, the control and enhancement of optical fields at the nanoscale is achieved by designing subwavelength resonators that can support localized Mie-like resonant modes. These optical modes enable the observation of third order nonlinear processes, (including third harmonic generation and four wave mixing) using centrosymmetric materials such as Si, [49–51] or Ge, [52,54] and second order nonlinear effects such as second harmonic generation in metasurfaces made from non-centrosymmetric materials such as GaAs, [33–39] AlGaAs, [42–48] GaP, [55] LiNbO3, [56] BatiO3 [57]. These harmonic generation processes were further enhanced via the engineering of high-quality factor, bound state in the continuum optical modes, [34,81] or in isolated individual resonators. [46] In all of these systems, it is paramount to have a good electric field overlap between the nonlinear tensor of the materials and the respective optical mode involved. In the following, we will describe our approach at engineering these two types of optical modes, namely, Mie-like resonances and Bound States in the Continuum (BICs).

At the Center for Integrated Nanotechnologies, we have been working with natural materials that possess high second order nonlinearities, such as GaAs, [84] and the respective ternary alloyed semiconductors AlGaAs. [85] To fabricate a metasurface we need to find ways of creating isolated nanoscale resonators. Since light tends to be confined to the material of higher refractive index, most strategies have relied on surrounding a volume of high index material in a lower dielectric medium. For example, silicon particles have been embedded in a lower index to create Huygens' metasurfaces, [86,87] or the chosen nonlinear material was transferred onto a glass substrate using various transfer techniques [15]. In our case, we choose to grow a multilayer GaAs/AlGaAs [100] substrate, pattern the grown material using electron beam lithography and chlorine-based, inductive coupled plasma etching, and then oxidize the non-exposed AlGaAs layers to AlGaO, a dielectric material with a lower refractive index ( $n_{GaAs} \approx 3.6; n_{AlGaO} \approx 1.6$) for near IR wavelengths. [33,88] The fabrication process is robust and reproducible. Figure 2a shows one example of such metasurface where we fabricated and characterized the second harmonic response; inset 2b illustrates the material layers described before. Figure 2c shows the respective orientation of the metasurface at 45degrees with respect to the crystal axis. [39] This approach constrains the experiments to be carried out in reflection mode as the GaAs substrate absorbs any harmonic light generated below 820 nm.

When shaped as nanocylinders, each optical antenna can support Mie-type electric and magnetic dipole modes (as well as higher order modes that won't be discussed here). The electric dipole mode is characterized by having the electric field mostly aligned with the incident electric field polarization as shown in the normalized distribution of the electric field intensity of figure 2d. For the magnetic dipole mode however, the electric field loops around a center axis of the GaAs particle's volume causing the optical mode to have both x and z electric field components. Due to the property of the nonlinear tensors of GaAs [33] and AlGaAs, [48] magnetic dipole excitation leads to stronger second harmonic generation (by two orders of magnitudes) compared to the electric dipole mode as shown in figure 2f.

The integral overlap between the nonlinear tensor of GaAs [100] and Mie-like resonant modes predicts little SHG from the metasurface to be emitted at normal incidence. Most of the SHG is emitted in an off-normal direction and diffracts into higher orders as shown by the radiation pattern collected at the back focal plane of the optical system (see the top two panels of figure 2f). [39] and requires higher numerical aperture collecting optics. In recent years, it has been proposed that a metasurface made of GaAs whose normal axis is aligned with the crystal axis [111] will emit most of its SHG at normal incidence. [38] In our case, when looking at the variation of the SHG as a function of the fundamental polarization, we find a fourfold symmetry intensity SHG signal characterized by minima at respective 45- and 135-degree polarization for electric dipole excitation, which corresponds to the situation where the incident electric field polarization is aligned with the respective crystal axis of GaAs. From the bulk symmetry tensor of the GaAs [100], [89,90] we expect to obtain minimum SHG for this polarization. For magnetic dipole mode excitation however, we find a twofold symmetry of the emitted SHG signal, that cannot be predicted from the nonlinear tensor of bulk GaAs alone. Twofold symmetry is characteristic of another crystal symmetry that can arise from tilted excitation [39] or from SHG originating from the surface of the resonator as also shown in the data of figure 2j. [33,40,89–92] Indeed, it has been predicted and observed that the surface of GaAs belongs to a symmetry point-group $mm2$ such that the non-zero elements of the nonlinear tensor obeys ($\chi^{(2,s)}_{zxx} = \chi^{(2,s)}_{zyy}, \chi^{(2,s)}_{zzz}, \chi^{(2,s)}_{yxz} = \chi^{(2,s)}_{xzy}$), which can lead to twofold symmetry of the SHG signal as a function of pump polarization. [40,89,90,93] Systematic modelling studies of surface SHG is challenging as it requires to find both an effective thickness in which surface effects are influential and choosing the right surface nonlinear coefficient. [40,91,92,94] Overall, these types of GaAs-based metasurfaces result in nonlinear extrinsic efficiencies on the order of $10^{-6}[W^{-1}]$. [33] In general, optical modes and nonlinear tensor symmetry can impact the generation of SHG. [95,96]

The relaxation of phase-matching also enables the observation of third harmonic signal generated via a cascaded process of sum frequency generation (SFG) and four wave mixing (FWM). [16,31,35] In general, given that both THG and its counterpart from cascaded generation occur at the same frequencies, and are assumed to have the same polarization response in cubic crystal systems, identifying the contribution of cascaded optical nonlinearities to the total nonlinear signal can be challenging. [45] In our latest work, we show that cascaded optical nonlinearities are in fact non-negligible in these systems, and are observable as a twofold symmetry in the total signal at $3\omega$ compared to the expected fourfold symmetry from $\chi^{(3)} - THG$. These findings suggest that alternative pathways to three wave mixing, and more broadly high harmonic generation can be achieved at the nanoscale, facilitated by the relaxation of phase-matching, thereby paving the way for a plethora of novel compact light conversion sources based on cascaded second order optical nonlinearities.

When using multiple incident pump beams, we can create a broadband dielectric optical mixer from these type of metasurfaces that can generate a multitude of new frequencies simultaneously facilitated by the relaxation of phase-matching conditions. [35] For example, as shown in figure 2j, when pumping our metasurface with two pump beams of respective wavelengths ($\lambda = 1240$ nm; $\lambda = 1570$ nm), we can generate respective, second (SHG), third (THG), and fourth (FHG) harmonic generation of each separate pump beams, and their mixing signals via four (FWM) or sixth wave mixing (SWM) as confirmed by the spectrogram of figure 2k where the frequency mixing signal that arises when the two pump beams overlap in time disappear for large time delay separation. These spectrogram data are instrumental for the characterization of ultrashort pulses since the shape, slope, and time elongation of each nonlinear trace gives information on the respective pulse widths, and chirps. [31] In GaAs, two photon absorption [97] can limit the efficiency of the nonlinear process. [33,35,40]

In conclusion, Mie- like resonant modes enhance the electromagnetic field within nonlinear nanoantennas which in turn boost the nonlinear processes such as harmonic generation over a broad range of operating wavelengths. [35] However, the quality factor of these resonances is low due to strong radiative coupling, which limits electromagnetic field enhancement and thus the nonlinear conversion efficiency. To bypass this limitation, we could use non-radiating anapole modes which have larger quality factors but a narrower operating bandwidth. [52] In general, there is a trade-off between quality factors and operating bandwidth. SHG can be enhanced using a doubly-resonant antenna approach where both optical modes are designed to respectively match the fundamental and harmonic frequencies. This method was vastly covered in plasmonic optical antennas and metasurfaces, [6,98,99] and recently demonstrated for semiconductor metasurfaces. [100]

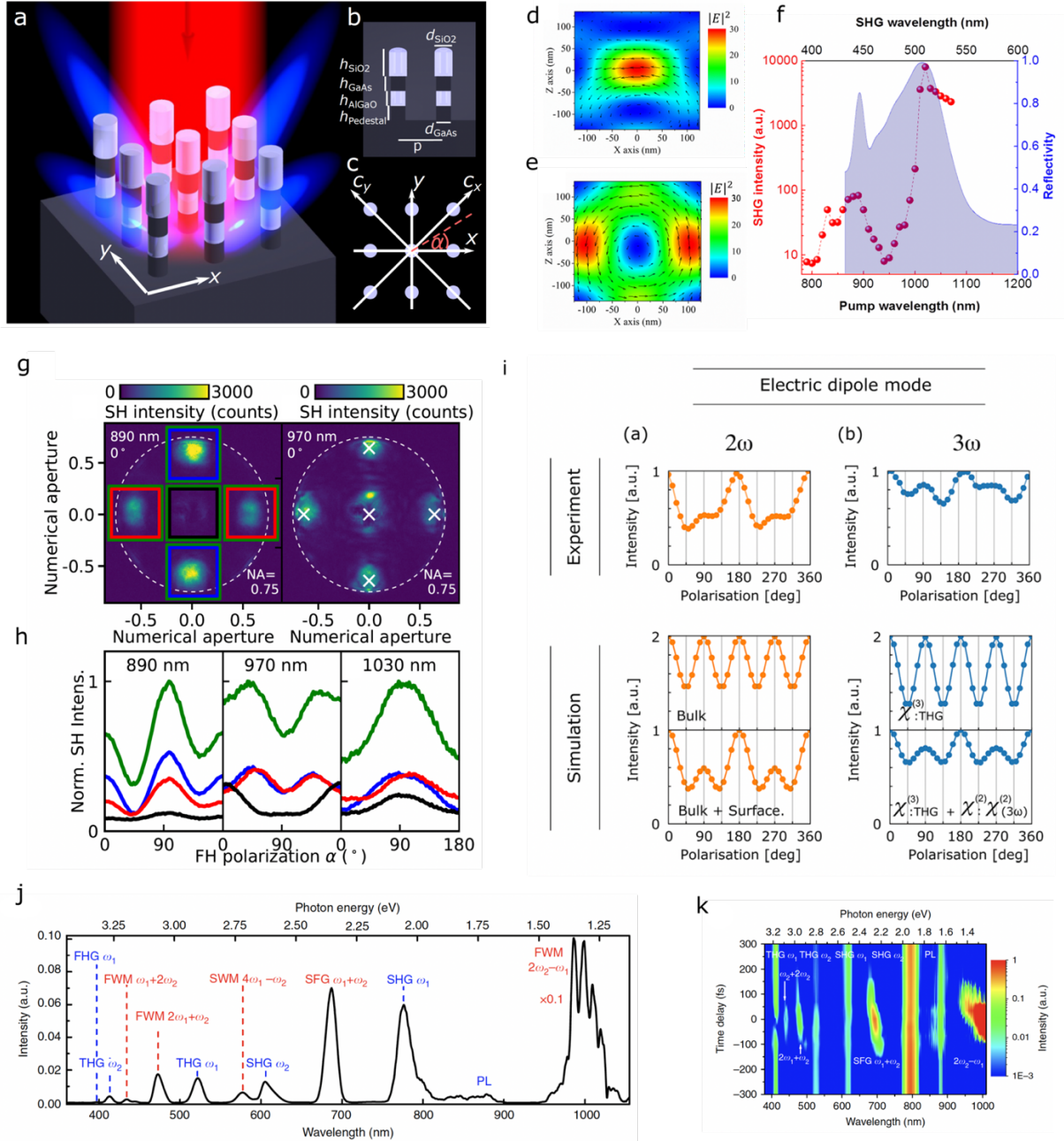

Figure 2 a) Artistic rendition of a GaAs nanocylinder metasurfaces excited by a pump (red) beam and generating second harmonic (blue) emitted via diffraction orders. (b) Cross-sectional sketch of nanocylinders showing the various material layers and geometric parameters. (c) Top view of the metasurfaces where the [010] and [001] crystal directions of the GaAs wafer are indicated by $c_x$ and $c_y$, with respect to the metasurface lattice directions x and y. Simulated normalized distribution of the electric intensity profile at the (d) electric and (e) magnetic dipole mode in the x–z plane cutting the GaAs nanoresonator in its middle. (f) Linear reflection spectra (filled blue) and second harmonic generation intensity (red dots) of a GaAs nanocylinder metasurfaces showing two peaks characteristic of a magnetic and an electric dipole mode. (g) Back focal plane imaging showing

SHG diffraction orders for pump wavelengths 890nm and 970nm. (h) SH intensity as a function of the incident fundamental polarization for three pump wavelengths. (i) Surface effect and cascaded optical nonlinearities manifest themselves as a twofold symmetry in the second and third harmonic polarization signal. (j) Generation of multiple harmonics via SHG, SFG, SWM, FWM, THG and FHG upon excitation of a GaAs nanocylinder metasurfaces with two pump beams at respective wavelengths of 1240nm and 1570nm. (k) Temporal dynamics of nonlinear frequency mixing. (j) Spectrogram in logarithmic scale when the time delay between the two pump pulses is varied. The shape of the trace can give information on the pulse duration and chirp. (k) Free carrier generation via two photon absorption and the induced resonant shift via refractive index change may cause a reduction of SHG when the two beams temporally overlap. Panels a, b, c, g and h are reprinted with permission from ref [39]; d, e, f from ref [33], i from ref [40] Copyright 2022. American Chemical Society. Panels j, k from ref [35] under a Creative Commons CC BY license.

Higher quality factor resonances enhance optical fields while increasing the efficiencies of harmonic generation processes. In recent years, there have been an increased interest in bound state in the continuum (BICs) optical modes in metasurfaces. [43,77,79,80,82,101,102] Proposed by von Neumann and Wigner in 1929, BICs are optical modes of discrete energy states that overlap within a continuous spectrum of radiating modes. [75–77,103] Symmetric-protected BICs are a type of BICs in which the coupling of the resonance to some radiative channels is forbidden by symmetry. [77,79] For example, in systems that exhibit a reflection or rotational symmetry, a bound state of one symmetry class can exist within a continuum spectrum of radiative states from another symmetry class, and their coupling is forbidden as long as the symmetry is preserved. So, the BIC has no radiative channels available to dissipate its energy and its quality factor becomes infinite. [82] Due to the finite size of most metasurfaces, most of these resonances are referred as 'quasi-BICs' with a finite high quality factor.

This type of symmetry-protected BIC has been explored in our research group back in 2016 where we considered an array of square resonator arranged in a square lattice though at the time we described their optical properties via a coupled mode 'Fano' interference, [34,75] in reference to the asymmetric lineshape observed in their plasmonic counterparts [104–108]. Their connection with BIC physics was later established in 2018 as symmetry-protected BICs. [77] Indeed, the whole metasurface obeys $C_2$ symmetry as shown in figure 3a. Each individual resonator can support orthogonal, but degenerate sets of electric and magnetic dipole modes along the x, y, and z directions. When these square resonators are arranged in an array with subwavelength periodicity, only the transverse (i.e., in-plane) dipole modes can couple to a normally incident electromagnetic wave and can scatter incident radiation. These optical modes manifesting themselves as broad peaks in the transmission and reflection resonant spectra. When breaking the symmetry of the system by adding a notch to the square resonator as illustrated by the artistic rendition of the metasurface in figure 3b, the non-radiating magnetic dipole mode $m_z$ of the system can now couple to an external field and a high-quality factor resonance can now be excited via far field incident radiation. It appears as a sharp asymmetric Fano lineshape in the transmission spectra due to the destructive interference with the scattering from the broad magnetic dipole modes present at these wavelengths. A similar process involving the radiative in-plane magnetic dipole ($m_y$) and non-radiative longitudinal electric dipole ($p_z$) leads to a second resonance with an asymmetric Fano lineshape at a higher energy.

Symmetry-protected BIC resonances with high quality factors have been used to enhance nonlinear processes at the nanoscale in metasurfaces.[34,75,77,81] Using the broken-symmetry design, we fabricated a metasurface in GaAs and characterized its second harmonic response when the fundamental beam excites the wavelength of the BIC resonance. We obtained a BIC resonance with a Q factor of 500 and observed up to threefold enhancement in SHG efficiency compared to traditional metasurfaces based on electric and magnetic dipoles though only 30% of the pulse energy could couple to the resonance due a bandwidth mismatch between the spectrum of the excitation pulse and the resonance.[34] Another type of BIC resonance can be formed where two optical modes couple to each other such that both modes exchange energy and one mode gets lossier whereas the other one gets less lossy such that its Q factor increases to infinity.[46,77] Named after the scientist who discovered them,[78] Friedrich-Wingten BIC has enabled the observation of strong second harmonic generation in a single antenna system.[46]

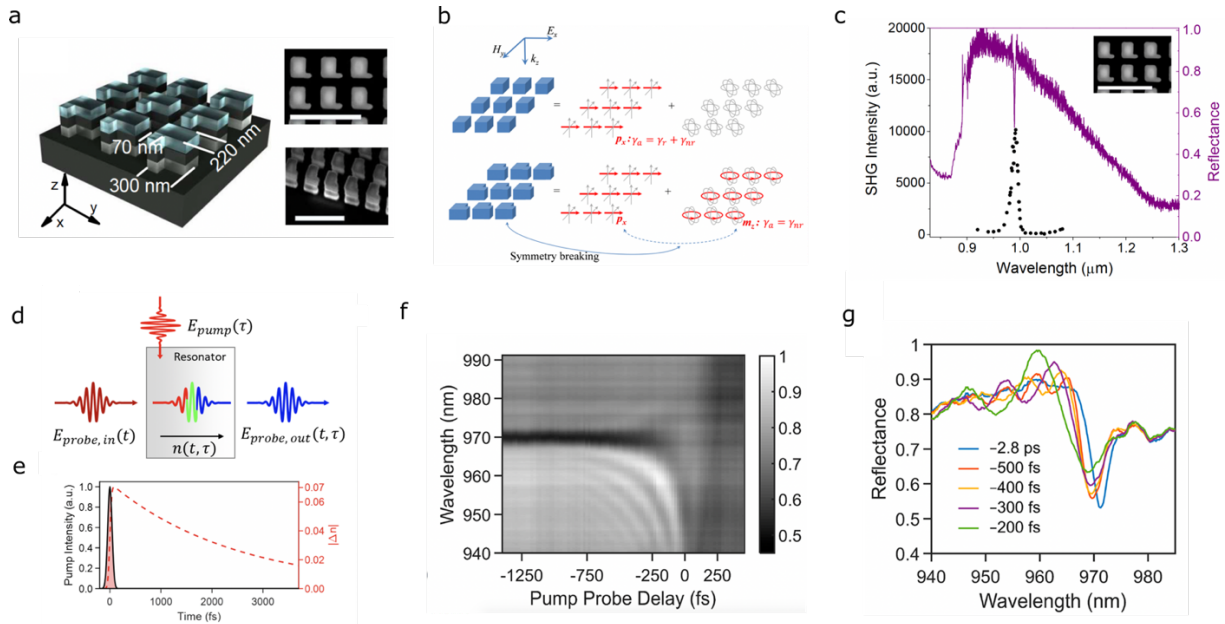

Figure 3. (a) Artistic rendition of the bound state in the continuum metasurface. Insets are scanning electron micrographs of the metasurface considered in the work.[34] (b) Schematic description of the operating principle of the BIC. While the incident electric field excites only the px electric dipole mode of the cube resonators (red arrows), symmetry breaking allow coupling of incident radiation to the out of plane magnetic dipole $m_z$. (c) Linear reflectance spectra (purple) and second harmonic generation (black dots) of the metasurface shown in (a). Enhanced field at the BIC resonance causes SHG from the metasurface to be dramatically increased. (d) Schematic illustrating frequency conversion in a time-variant resonator. An input pulse (red) couples into an optical mode of a resonator, which simultaneously undergoes a refractive index shift due to an external pump. This blueshifts the mode resonance causing new frequencies to be added in the output pulse (blue). (e) Simulated refractive index shift (red dashed) of a GaAs resonator versus time under a short (80 fs) pump pulse (red area). The index shifts within ~100fs and then decays according to free carrier relaxation over several picoseconds. (f) Experimental transient reflectance spectra showing interference fringes from the frequency converted light and the portion of the probe that is not resonant. (f) Individual spectra of the reflectance from the metasurface at different vertical cross section (i.e. pump delay) of (e). Panels a, c, are reprinted with permission from ref [34], panels b from ref [75] and panels d, e, f, g from [109]. Copyright 2022. American Chemical Society.

Nonlinear quasi-BIC metasurfaces are also attractive for ultrafast tunable metasurfaces within picosecond response times. [41,109–112] Since electronic devices are limited in bandwidths due to their slow temporal response typically limited by RC response times, all-optical approaches have been explored as an alternative route toward faster signal modulators and processors. [111] In these systems, a strong optical pulse induces a transient change of the optical properties of a system via single or multiphoton absorption causing free carriers to be generated and relax through various scattering processes including electron-phonon relaxation as illustrated in figure 3d. Theses phonons then dissipate their energy via heat transfer to their surrounding environments. Free carrier generation causes a noticeable refractive index that causes a metasurface resonance to change its bandwidth and shift in wavelength (see Fig. 3e). This effect generally manifests in metasurfaces as a change of reflectivity, absorptivity or transmittivity of a temporally-overlapping second pulse, [109,111] or as a transient change of the second harmonic generation under above-bandgap pump excitation [41,47].

When combining time-variant materials with a BIC metasurface that supports high-quality factor resonances, we demonstrate in our work how a rapidly shifting refractive index due to an external optical pump can induce frequency conversion of a second light pulse coupled inside the metasurface nanoresonators. [109] As shown in the spectrogram of figure 3d, and the respective time-delay cross section of figure 3e, the frequency conversion appears as interference fringes between the frequency converted light pulses and the light pulse unhindered by the metasurface BIC resonances. The intersection of high quality-factor resonances, active materials, and ultrafast transient spectroscopy enables metasurfaces operating in a time-variant regime demonstrating additional route toward frequency conversion. Other notable approaches to enhancing nonlinear harmonic generation at the nanoscale have been the use of topology to embed electric field into defect-protected edge, or corner states of photonics crystals and metasurfaces, [50,113,114] .

As a final remark, it is also worth mentioning that the ultrafast optical excitation of second order optical nonlinear materials can also lead to the generation of bursts of terahertz (THz) radiation via (optical) rectification and shift currents. [115–117] In particular, semiconductor metasurfaces and optical antennas have been used in recent years to both enhance the generation, [94,118] and detection [119–121] of terahertz radiation.

# Frequency Conversion via Resonant Nonlinearities in All-Dielectric Metasurfaces

While the approaches for nonlinear wave mixing described in the previous section using virtual transitions in conventional III-V semiconductors have led to numerous fundamentally interesting studies at CINT, a factor that limits the efficiencies of these nonlinear frequency mixing systems is the magnitude of each individual tensor element, that constitute a conventional material's nonlinear susceptibility. These values are set by nature. Therefore, efficient harmonic generation in these material systems typically require high pump fluences. This increases the conversion efficiency of the frequency mixing process under study, but also that of unwanted nonlinear processes such as two photon absorption and free carrier generation, which now compete on an equal footing for the incident pump beam energy. Such effect manifests as a saturation of the nonlinear signal for increasing pump powers as shown in our previous works [33,40]. Moreover, high pump fluences are generally achieved via expensive, energy-intensive, mode-locked pulse laser systems. Boosting the magnitude of the nonlinear susceptibility of the material becomes even more paramount when we now factor the cost and system-compactness of the associated excitation sources. If only, one could engineer the nonlinear susceptibilities of a material at will, this would give us more flexibility in designing compact nonlinear systems.

In recent years, quantum-engineered intersubband transitions supported by III-V semiconductor heterostructures have been a promising alternative to create materials with giant second order nonlinearities. [65,66,122] Unlike virtual transitions in bulk III-V semiconductors, resonant nonlinearities associated with intersubband transitions (ISTs) involve real electronic transitions, which, when properly engineered, can generate giant nonlinearities on the order of ∼ 200 nm/V, [67] which is 3 orders of magnitude larger than what can be achieved using conventional materials such as GaAs. [84] Figure 4(a) shows a band structure calculation of a typical III-V semiconductor heterostructure that we have used in our work [59]. It consists of two asymmetric coupled InGaAs quantum wells with AlInAs barriers. These quantum wells are grown on a lattice matched InP substrate via molecular beam epitaxy. The thicknesses of these quantum wells and the barriers are optimized such that they support three electronic levels that are equally spaced in energy with the largest possible transition dipole moments. The asymmetry along z in the design is required for second-harmonic generation to break the inversion symmetry. Furthermore, due to IST selection rules, only polarized electric fields normal to the quantum wells can couple to the ISTs. Therefore, the second order nonlinear susceptibility tensor of such system only has a non-zero element, $\chi^{(2)}_{zzz}$, which can be approximated by [59]:

$$\chi^{(2)} = \frac{e^3}{\epsilon_0 \hbar^2} \frac{N_d z_{12} z_{23} z_{31}}{(\omega - \omega_{12} - i\gamma_{12})(2\omega - \omega_{13} - i\gamma_{13})} \qquad (2)$$

where $e$ is the elementary charge, $\epsilon_0$ is the vacuum permittivity, $\hbar$ is the reduced Plank constant, $N_d$ is the average bulk doping density, $\gamma_{ij}$ represents the IST damping rate, and $z_{ij}$ are the transition dipole matrix lengths between the energy levels *i* and *j* of the quantum wells. The calculated $\chi^{(2)}_{zzz}$, corresponding to the energy band structure shown in Fig. 4(a), is shown in Fig. 4(b). In this case, the ISTs were designed such that the $\chi^{(2)}_{zzz}$ peaks at approximately 250 nm/V for SHG at a pump wavelength of 10 μm, which is orders of magnitude larger than bulk $\chi^{(2)}$ of standard III-V semiconductors. Working with IST in III-V semiconductor heterostructures thus require electric fields with polarization normal to a materials surface, which is generally hard to achieve with conventional free space optics. Instead, waveguide geometries and plasmonic metasurfaces are used to generate strong z-field components. [60] However, real transitions have associated absorption losses and are also prone to saturation effects [59]. It is thus paramount to design low-powered photonics systems to limit these impacts.

At CINT, we have incorporated all-dielectric metasurfaces with photonic modes coupled to IST quantum wells (QWs) for efficient second harmonic generation. [58,59,67] Our first approach is to design a dielectric metasurface consisting of a germanium grating with a guiding layer underneath fabricated on top of a multi-QW layers as shown in figure 4c. The structure supports leaky mode resonances at both pump and second-harmonic frequencies, which couple to the IST via their z-polarized, evanescent fields. [58] Thus, the metasurface enables both efficient in-coupling of the incident radiation to the ISTs, and outcoupling of the generated second-harmonic signal, thereby achieving strong second-harmonic generation. Furthermore, these leaky mode resonances can be spectrally tuned by tilting the metasurface with respect to incident radiation, which tune the second harmonic bandwidth of the system. Thus, we can operate the metasurface over a broad range of wavelengths (See Fig. 4. d).

Leaky-mode resonances do have some drawbacks though. First, they cannot be excited with normal incident radiation. Secondly, though the field inside these modes can be quite intense, only a fraction of it couples to the IST via evanescent fields, and the field intensity decreases as the distance from the grating/IST interface increases. In other words, the nonlinear medium (IST quantum well) is not placed at the location where the field is the strongest, thus limiting the nonlinear efficiency of the whole system. In a subsequent work, we also show how stronger light-matter coupling can be achieved when the quality factors of both the material and photonics resonances are similar.

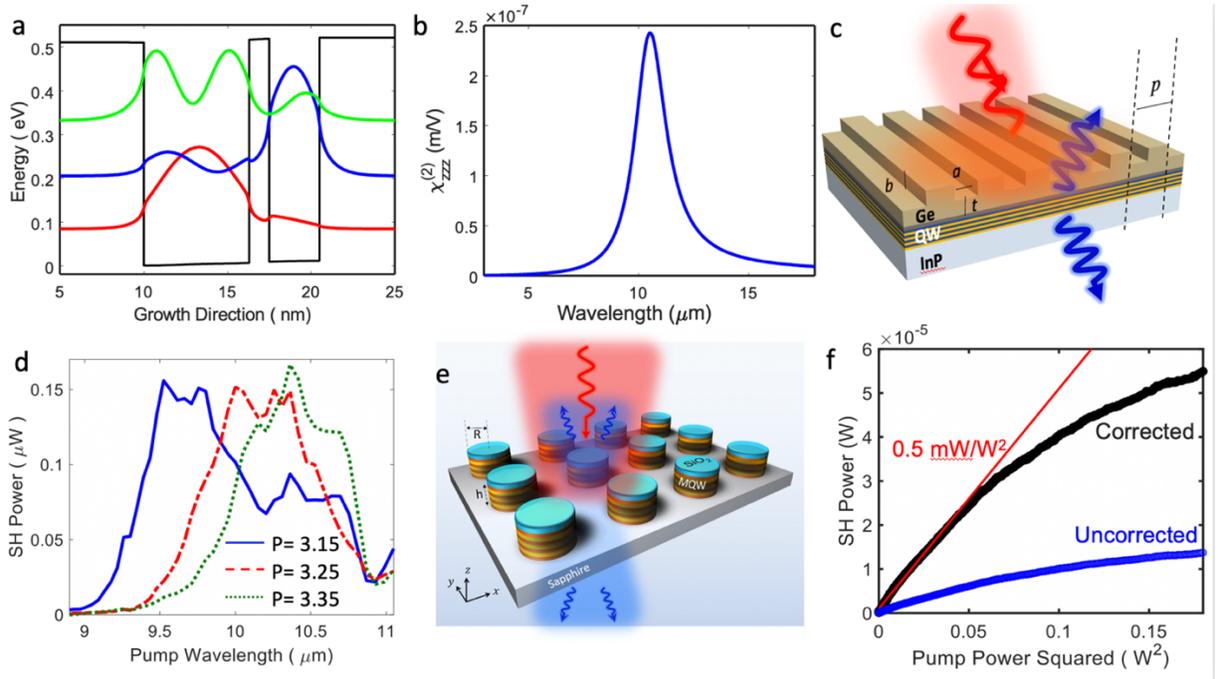

Figure 4: (a) 8 band k.p band structure calculation of a single period of a doubly resonant asymmetric coupled quantum-well structure. The thicknesses of the layers (AlInAs/InGaAs/AlInAs/InGaAs/AlInAs) are 10 nm/6.3 nm/1.2 nm/3 nm/10 nm. (b) Intersubband nonlinear susceptibility of the multi-QW structure shown in (a) as a function of pump wavelength calculated using equation 1. (c) Schematic of the nonlinear metasurface that supports leaky mode resonances. The geometric parameters of the fabricated devices are a = 1070 nm, b= 1200 nm, t= 500 nm, and p = 3150, 3250, 3350 nm (for different structures measured). The thickness of the multi-QW layer is 656 nm. (d) Reflected SH power as a function of pump wavelength from structures with p = 3.15 $\mu$m (blue solid line), 3.25 $\mu$m (red dashed line), and 3.35 $\mu$m (green dotted line) at an incident pump intensity of approximately 3.3 kW/cm$^2$ (e) Schematic of the nonlinear metasurface comprised of cylindrical meta-atoms with multi-QWs embedded within. The height ($h$) of all the fabricated cylindrical Mie resonators is 1.5 $\mu$m. The radii ($R$) of the cylinders are varied to vary the spectral position of the Mie photonic resonances in order to couple incident pump light to the intersubband transitions (f) Experimentally-measured peak SH power as function of square of incident peak pump power at pump wavelength of 7.65 $\mu$m. The figure includes the raw data (uncorrected) as well as the data corrected for the limited collection efficiency at 7.65 $\mu$m. The slope of the linear fit (red line) at lower pump powers determines the nonlinear conversion factor $\eta$ which is equal to 0.5 mW/W$^2$. The deviation of the slope at higher pump powers is seen due to the saturation effect of the ISTs. Panels a, b, e and f are reprinted with permission from ref [67], panels c and d from ref. [58]. Copyright 2022 American Chemical Society.

All these limitations can be overcome by shaping the IST quantum wells as nanoscale resonators as shown in figure 4e such that the Mie-type photonic modes that exists in them spectrally overlap with the IST transitions. Moreover, magnetic-dipole modes have a strong z-field components (see the field distribution of figure 2e), which is ideal to excite ISTs. Moreover, the linewidths of these photonic resonances can be made to match that of the ISTs and achieve strong-light matter coupling [59] which allows us to achieve efficient second harmonic generation. As a case in point, we design and fabricated a metasurface that can operate at normal and off-normal incidence where the nanoscale resonator had optimized multi-QWs embedded within them (figure 4e). This enables us to efficiently couple the pump light via magnetic dipole excitation and achieve the highest second harmonic generation efficiency reported to date using all-dielectric metasurfaces (see results shown in Fig. 4f).

# Conclusion

In conclusion, we have reviewed our recent work on nonlinear and ultrafast GaAs-based metasurfaces and highlighted various approaches based on field enhancement to promote frequency conversion. Since harmonic generation processes scale nonlinearly with the electric field, we have shown the importance of engineering Mie-like resonance and high-quality factor BIC optical modes to boost the optical fields within nanoscale resonators. When paired with a second pump beam, we have demonstrated that transient changes in the material's refractive index of a metasurface with BIC resonances causes new frequencies to appear. While nonlinear metasurfaces for harmonic generation have been explored in great lengths over the past few years, recent work on cascaded optical nonlinearities in semiconductor metasurfaces [40] may open up new avenues to engineer a variety of novel optical phenomena such as all-optical modulators, transistor action, mode-locking via $\chi^{(2)} - \chi^{(2)}$ nonlinearities, and spatial soliton waves within metasurfaces. [123,124] Another future research direction is on transient harmonic generation where a pump optical beam dynamically changes the optical property of a material to modify and modulate the strength and momenta of the harmonic processes. [41,47]

In addition to above, we have also reviewed our recent work on all-dielectric nonlinear polaritonic metasurfaces where we couple photonic resonances to matter excitations such as ISTs to generate giant nonlinear responses. With metasurfaces where the IST are embedded inside Mie resonators, we have demonstrated the highest second harmonic conversion efficiency reported to date. There is still room for improvement in such metasurfaces by optimizing the interplay between the field enhancements of the photonic modes and the material nonlinearity, while minimizing the associated absorption due to real transitions, and controlling the harmonic photons' emission directionality. This research is currently undergoing at CINT. Furthermore, since we have used InGaAs/AlInAs heterostructures, our metasurfaces have operated in the mid-infrared wavelength regime. In future, an exciting research direction will be to investigate other material systems for all-dielectric nonlinear metasurfaces such as III-Sb and III-Nitrides to bring the wavelength of operation to near-IR wavelengths and provide the near-IR regime with nonlinear efficiencies never dreamed of. [62,125]

For the future, we envision the field of nonlinear metasurfaces going toward the control of quantum optical nonlinearities with various potential application as outlined in this recent review. [126] Bound states in the continuum metasurfaces have gained momentum in recent years as enhanced optical fields could boost optical nonlinearities such as harmonic generation, but also their reverse process of spontaneous parametric down conversion and spontaneous four wave mixing. While photon pairs generation have been observed in thin $LiNBO_3$ substrates, [127] optical antennas, [128] and metasurfaces, [129] it has been theoretically predicted that bound state in the continuum modes could drastically enhance the rate of pair generation at the nanoscale, and help create novel quantum sources. [130] We also envision that the combination of bound states in the continuum with resonant nonlinearities can open up interesting physics questions encompassing quantum optics and polaritonic nanophotonic systems.

**Funding.** U.S. Department of Energy (BES 20-017574);

**Acknowledgments.** This work was supported by the U.S. Department of Energy, Office of Basic Energy Sciences, Division of Materials Sciences and Engineering and performed, in part, at the Center for Integrated Nanotechnologies, an Office of Science User Facility operated for the U.S. Department of Energy (DOE) Office of Science. Sandia National Laboratories is a multi-mission laboratory managed and operated by National Technology and Engineering Solutions of Sandia, LLC, a wholly owned subsidiary of Honeywell International, Inc., for the U.S. Department of Energy's National Nuclear Security Administration under contract DE-NA0003525. This paper describes objective technical results and analysis. Any subjective views or opinions that might be expressed in the paper do not necessarily represent the views of the U.S. Department of Energy or the United States Government.

**Disclosures.** The authors declare no conflict of interest

**References**

1.  T. H. MAIMAN, "Stimulated Optical Radiation in Ruby," Nature **187**, 493–494 (1960).

2.  R. W. Boyd, *Nonlinear Optics* (ACADEMIC PRESS, 2003).

3.  P. A. Franken, A. E. Hill, C. W. Peters, and G. Weinreich, "Generation of Optical Harmonics," Phys. Rev. Lett. **7**, 118–119 (1961).

4.  M. H. Chou, I. Brener, M. M. Fejer, E. E. Chaban, and S. B. Christman, "1.5-μm-band wavelength conversion based on cascaded second-order nonlinearity in LiNbO 3 waveguides," IEEE Photonics Technol. Lett. **11**, 653–655 (1999).

5.  M. P. Nielsen, X. Shi, P. Dichtl, S. A. Maier, and R. F. Oulton, "Giant nonlinear response at a plasmonic nanofocus drives efficient four-wave mixing," Science **358**, 1179–1181 (2017).

6.  S. D. Gennaro, M. Rahmani, V. Giannini, H. Aouani, T. P. H. H. Sidiropoulos, M. Navarro-Cía, S. A. Maier, and R. F. Oulton, "The Interplay of Symmetry and Scattering Phase in Second Harmonic Generation from Gold Nanoantennas," Nano Lett. **16**, 5278–5285 (2016).

7.  A. Krasnok, M. Tymchenko, and A. Alù, "Nonlinear metasurfaces: a paradigm shift in nonlinear optics," Mater. Today **21**, 8–21 (2018).

8.  D. K. Gramotnev and S. I. Bozhevolnyi, "Nanofocusing of electromagnetic radiation," Nat. Photonics **8**, 13–22 (2013).

9.  L. Novotny and N. F. Van Hulst, "Antennas for light," Nat. Photonics **5**, 83–90 (2011).

10. J. A. Schuller, E. S. Barnard, W. Cai, Y. C. Jun, J. S. White, and M. L. Brongersma, "Plasmonics for extreme light concentration and manipulation.," Nat. Mater. **9**, 193–204 (2010).

11. L. Bonacina, P.-F. Brevet, M. Finazzi, and M. Celebrano, "Harmonic generation at the nanoscale," J. Appl. Phys. **127**, 2309011–23090117 (2020).

12. L. Lafone, T. P. H. Sidiropoulos, and R. F. Oulton, "Silicon-based metal-loaded plasmonic waveguides for low-loss nanofocusing," Opt. Lett. **39**, 4356–4359 (2014).

13. D. Pohl, M. Reig Escalé, M. Madi, F. Kaufmann, P. Brotzer, A. Sergeyev, B. Guldimann, P. Giaccari, E. Alberti, U. Meier, and R. Grange, "An integrated broadband spectrometer on thin-film lithium niobate," Nat. Photonics **14**, 24–29 (2020).

14. G. Li, S. Zhang, and T. Zentgraf, "Nonlinear photonic metasurfaces," Nat. Rev. Mater. **2**, 1701011–1701014 (2017).

15. M. Rahmani, G. Leo, I. Brener, A. V. Zayats, S. A. Maier, C. De Angelis, H. H. Tan, V. Flavio Gili, Fouad Karouta, R. F. Oulton, K. Vora, M. Lysevych, Isabelle Staude, L. Xu, A. E. Miroshnichenko, C. Jagadish, and D. N. Neshev, "Nonlinear frequency conversion in optical nanoantennas and metasurfaces: materials evolution and fabrication," Opto-Electronic Adv. **1**, 18002101–18002112 (2018).

16. M. Celebrano, A. Locatelli, L. Ghirardini, G. Pellegrini, P. Biagioni, A. Zilli, X. Wu, S. Grossmann, L. Carletti, C. De Angelis, L. Duò, B. Hecht, and M. Finazzi, "Evidence of Cascaded Third-Harmonic Generation in Noncentrosymmetric Gold Nanoantennas," Nano Lett. **19**, 7013–7020 (2019).

17. B. Sain, C. Meier, and T. Zentgraf, "Nonlinear optics in all-dielectric nanoantennas and metasurfaces: a review," Adv. Photonics **1**, 0240021–02400214 (2019).

18. M. R. Shcherbakov, F. Eilenberger, and I. Staude, "Interaction of semiconductor metasurfaces with short laser pulses: From nonlinear-optical response toward spatiotemporal shaping," J. Appl. Phys. **126**, 0857051–08570513 (2019).

19. S. D. Gennaro, T. R. Roschuk, S. A. Maier, and R. F. Oulton, "Measuring chromatic aberrations in imaging systems using plasmonic nanoparticles," Opt. Lett. **41**, 1688–1691 (2016).

20. M. W. Knight, N. S. King, L. Liu, H. O. Everitt, P. Nordlander, and N. J. Halas, "Aluminum for Plasmonics," ACS Nano **8**, 834–840 (2014).

21. S. A. Maier, *Plasmonics : Fundamentals and Applications* (Springer US, 2007).

22. M. Celebrano, X. Wu, M. Baselli, S. Großmann, P. Biagioni, A. Locatelli, C. De Angelis, G. Cerullo, R. Osellame, B. Hecht, L. Duò, F. Ciccacci, and M. Finazzi, "Mode matching in multiresonant plasmonic nanoantennas for enhanced second harmonic generation," Nat. Nanotechnol. **10**, 412–417 (2015).

23. H. Aouani, M. Navarro-cia, M. Rahmani, T. P. H. Sidiropoulos, M. Hong, R. F. Oulton, and S. A. Maier, "Multiresonant Broadband Optical Antennas As Efficient Tunable Nanosources of Second Harmonic Light," Nano Lett. **12**, 4997–5002 (2012).

24. S. Kujala, B. K. Canfield, M. Kauranen, Y. Svirko, and J. Turunen, "Multipole Interference in the Second-Harmonic Optical Radiation from Gold Nanoparticles," Phys. Rev. Lett. **98**, 1674031–1674034 (2007).

25. J. Butet, P.-F. Brevet, and O. J. F. Martin, "Optical Second Harmonic Generation in Plasmonic Nanostructures: From Fundamental Principles to Advanced Applications," ACS Nano **9**, 10545–10562 (2015).

26. K. O'Brien, H. Suchowski, J. Rho, A. Salandrino, B. Kante, X. Yin, and X. Zhang, "Predicting nonlinear properties of metamaterials from the linear response," Nat. Mater. **14**, 379–383 (2015).

27. M. Kauranen and A. V. Zayats, "Nonlinear plasmonics," Nat. Photonics **6**, 737–748 (2012).

28. G. Sartorello, N. Olivier, J. Zhang, W. Yue, D. J. Gosztola, G. P. Wiederrecht, G. Wurtz, and A. V. Zayats, "Ultrafast Optical Modulation of Second- and Third-Harmonic Generation from Cut-Disk-Based Metasurfaces," ACS Photonics **3**, 1517–1522 (2016).

29. B. Metzger, L. Gui, J. Fuchs, D. Floess, M. Hentschel, and H. Giessen, "Strong Enhancement of Second Harmonic Emission by Plasmonic Resonances at the Second Harmonic Wavelength," Nano Lett. **15**, 3917–3922 (2015).

30. S. D. Gennaro, Y. Li, S. A. Maier, and R. F. Oulton, "Nonlinear Pancharatnam–Berry Phase Metasurfaces beyond the Dipole Approximation," ACS Photonics **6**, 2335–2341 (2019).

31. S. D. Gennaro, Y. Li, S. A. Maier, and R. F. Oulton, "Double Blind Ultrafast Pulse Characterization by Mixed Frequency Generation in a Gold Antenna," ACS Photonics **5**, 3166–3171 (2018).

32. H. Aouani, M. Rahmani, M. Navarro-Cía, and S. A. Maier, "Third-harmonic-upconversion enhancement from a single semiconductor nanoparticle coupled to a plasmonic antenna," Nat. Nanotechnol. **9**, 290–294 (2014).

33. S. Liu, M. B. Sinclair, S. Saravi, G. A. Keeler, Y. Yang, J. Reno, G. M. Peake, F. Setzpfandt, I. Staude, T. Pertsch, and I. Brener, "Resonantly Enhanced Second-Harmonic Generation Using III-V Semiconductor All-Dielectric Metasurfaces," Nano Lett. **16**, 5426–5432 (2016).

34. P. P. Vabishchevich, S. Liu, M. B. Sinclair, G. A. Keeler, G. M. Peake, and I. Brener, "Enhanced Second-Harmonic Generation Using Broken Symmetry III-V Semiconductor Fano Metasurfaces," ACS Photonics **5**, 1685–1690 (2018).

35. S. Liu, P. P. Vabishchevich, A. Vaskin, J. L. Reno, G. A. Keeler, M. B. Sinclair, I. Staude, and I. Brener, "An all-dielectric metasurface as a broadband optical frequency mixer," Nat. Commun. **9**, 2507 (2018).

36. M. Timofeeva, L. Lang, F. Timpu, C. Renaut, A. Bouravleuv, I. Shtrom, G. Cirlin, and R. Grange, "Anapoles in Free-Standing III-V Nanodisks Enhancing Second-Harmonic Generation," Nano Lett. **18**, 3695–3702 (2018).

37. L. Xu, G. Saerens, M. Timofeeva, D. A. Smirnova, I. Volkovskaya, M. Lysevych, R. Camacho-Morales, M. Cai, K. Zangeneh Kamali, L. Huang, F. Karouta, H. H. Tan, C. Jagadish, A. E. Miroshnichenko, R. Grange, D. N. Neshev, and M. Rahmani, "Forward and Backward Switching of Nonlinear Unidirectional Emission from GaAs Nanoantennas," ACS Nano **14**, 1379–1389 (2020).

38. J. D. Sautter, L. Xu, A. E. Miroshnichenko, M. Lysevych, I. Volkovskaya, D. A. Smirnova, R. Camacho-Morales, K. Zangeneh Kamali, F. Karouta, K. Vora, H. H. Tan, M. Kauranen, I. Staude, C. Jagadish, D. N. Neshev, and M. Rahmani, "Tailoring Second-Harmonic Emission from (111)-GaAs Nanoantennas," Nano Lett. **19**, 3905–3911 (2019).

39. F. J. F. Löchner, A. N. Fedotova, S. Liu, G. A. Keeler, G. M. Peake, S. Saravi, M. R. Shcherbakov, S. Burger, A. A. Fedyanin, I. Brener, T. Pertsch, F. Setzpfandt, and I. Staude, "Polarization-Dependent Second Harmonic Diffraction from Resonant GaAs Metasurfaces," ACS Photonics **5**, 1786–1793 (2018).

40. S. D. Gennaro, C. F. Doiron, N. Karl, P. P. Iyer, D. K. Serkland, M. B. Sinclair, and I. Brener, "Cascaded Optical Nonlinearities in Dielectric Metasurfaces," ACS Photonics **9**, 1026–1032 (2022).

41. P. P. Vabishchevich, A. Vaskin, N. Karl, J. L. Reno, M. B. Sinclair, I. Staude, and I. Brener, "Ultrafast all-optical diffraction switching using semiconductor metasurfaces," Appl. Phys. Lett. **118**, 2111051–2111056 (2021).

42. V. F. Gili, L. Carletti, A. Locatelli, D. Rocco, M. Finazzi, L. Ghirardini, I. Favero, C. Gomez, A. Lemaître, M. Celebrano, C. De Angelis, and G. Leo, "Monolithic AlGaAs second-harmonic nanoantennas," Opt. Express **24**, 15965–15971 (2016).

43. L. Carletti, S. S. Kruk, A. A. Bogdanov, C. De Angelis, and Y. Kivshar, "High-harmonic generation at the nanoscale boosted by bound states in the continuum," Phys. Rev. Res. **1**, 0230161–02301617 (2019).

44. L. Carletti, A. Locatelli, D. Neshev, and C. De Angelis, "Shaping the Radiation Pattern of Second-Harmonic Generation

from AlGaAs Dielectric Nanoantennas," ACS Photonics **3**, 1500–1507 (2016).

45. A. Zilli, D. Rocco, M. Finazzi, A. Di Francescantonio, L. Duò, C. Gigli, G. Marino, G. Leo, C. De Angelis, and M. Celebrano, "Frequency Tripling via Sum-Frequency Generation at the Nanoscale," ACS Photonics **8**, 1175–1182 (2021).

46. K. Koshelev, S. Kruk, E. Melik-Gaykazyan, J.-H. H. Choi, A. Bogdanov, H.-G. G. Park, and Y. Kivshar, "Subwavelength dielectric resonators for nonlinear nanophotonics," Science **367**, 288–292 (2020).

47. E. A. A. Pogna, M. Celebrano, A. Mazzanti, L. Ghirardini, L. Carletti, G. Marino, A. Schirato, D. Viola, P. Laporta, C. De Angelis, G. Leo, G. Cerullo, M. Finazzi, and G. Della Valle, "Ultrafast, All Optically Reconfigurable, Nonlinear Nanoantenna," ACS Nano **15**, 11150–11157 (2021).

48. L. Carletti, A. Locatelli, O. Stepanenko, G. Leo, and C. De Angelis, "Enhanced second-harmonic generation from magnetic resonance in AlGaAs nanoantennas," Opt. Express **23**, 265441 (2015).

49. M. R. Shcherbakov, D. N. Neshev, B. Hopkins, A. S. Shorokhov, I. Staude, E. V. Melik-Gaykazyan, M. Decker, A. A. Ezhov, A. E. Miroshnichenko, I. Brener, A. A. Fedyanin, and Y. S. Kivshar, "Enhanced third-harmonic generation in silicon nanoparticles driven by magnetic response," Nano Lett. **14**, 6488–6492 (2014).

50. S. Kruk, A. Poddubny, D. Smirnova, L. Wang, A. Slobozhanyuk, A. Shorokhov, I. Kravchenko, B. Luther-Davies, and Y. Kivshar, "Nonlinear light generation in topological nanostructures," Nat. Nanotechnol. **14**, 126–130 (2019).

51. T. Shibanuma, G. Grinblat, P. Albella, and S. A. Maier, "Efficient Third Harmonic Generation from Metal-Dielectric Hybrid Nanoantennas," Nano Lett. **17**, 2647–2651 (2017).

52. G. Grinblat, Y. Li, M. P. Nielsen, R. F. Oulton, and S. A. Maier, "Enhanced third harmonic generation in single germanium nanodisks excited at the anapole mode," Nano Lett. **16**, 4635–4640 (2016).

53. S. D. Gennaro, M. Goldflam, D. B. Burckel, J. Jeong, M. B. Sinclair, and I. Brener, "Dielectric metasurfaces made from vertically oriented nanoresonators," J. Opt. Soc. Am. B **38**, C33–C41 (2021).

54. G. Grinblat, Y. Li, M. P. Nielsen, R. F. Oulton, and S. A. Maier, "Efficient Third Harmonic Generation and Nonlinear Subwavelength Imaging at a Higher-Order Anapole Mode in a Single Germanium Nanodisk," ACS Nano **11**, 953–960 (2017).

55. J. Cambiasso, G. Grinblat, Y. Li, A. Rakovich, E. Cortés, and S. A. Maier, "Bridging the Gap between Dielectric Nanophotonics and the Visible Regime with Effectively Lossless Gallium Phosphide Antennas," Nano Lett. **17**, 1219–1225 (2017).

56. F. Timpu, J. Sendra, C. Renaut, L. Lang, M. Timofeeva, M. T. Buscaglia, V. Buscaglia, and R. Grange, "Lithium Niobate Nanocubes as Linear and Nonlinear Ultraviolet Mie Resonators," ACS Photonics **6**, 545–552 (2019).

57. E. Kim, A. Steinbrück, M. T. Buscaglia, V. Buscaglia, T. Pertsch, R. Grange, A. Steinbru, M. T. Buscaglia, V. Buscaglia, T. Pertsch, R. Grange, and K. I. M. E. T. Al, "Second-Harmonic Generation of Single Diameter," ACS Nano **7**, 5343–5349 (2013).

58. R. Sarma, D. De Ceglia, N. Nookala, M. A. Vincenti, S. Campione, O. Wolf, M. Scalora, M. B. Sinclair, M. A. Belkin, and I. Brener, "Broadband and Efficient Second-Harmonic Generation from a Hybrid Dielectric Metasurface/Semiconductor Quantum-Well Structure," ACS Photonics **6**, 1458–1465 (2019).

59. R. Sarma, N. Nookala, K. J. Reilly, S. Liu, D. De Ceglia, L. Carletti, M. D. Goldflam, S. Campione, K. Sapkota, H. Green, G. T. Wang, J. Klem, M. B. Sinclair, M. A. Belkin, and I. Brener, "Strong Coupling in All-Dielectric Intersubband Polaritonic Metasurfaces," Nano Lett. **21**, 367–374 (2021).

60. J. Lee, M. Tymchenko, C. Argyropoulos, P.-Y. Chen, F. Lu, F. Demmerle, G. Boehm, M. A. Belkin, M. Amann, A. Alù, and M. A. Belkin, "Giant nonlinear response from plasmonic metasurfaces coupled to intersubband transitions.," Nature **511**, 65–69 (2014).

61. J. Lee, N. Nookala, J. S. Gomez-Diaz, M. Tymchenko, F. Demmerle, G. Boehm, M. C. Amann, A. Alù, and M. A. Belkin, "Ultrathin Second-Harmonic Metasurfaces with Record-High Nonlinear Optical Response," Adv. Opt. Mater. **4**, 664–670 (2016).

62. O. Wolf, A. A. Allerman, X. Ma, J. R. Wendt, A. Y. Song, E. A. Shaner, and I. Brener, "Enhanced optical nonlinearities in the near-infrared using III-nitride heterostructures coupled to metamaterials," Appl. Phys. Lett. **107**, 1511081–1511085 (2015).

63. N. Nookala, J. Xu, O. Wolf, S. March, R. Sarma, S. Bank, J. Klem, I. Brener, and M. Belkin, "Mid-infrared second-harmonic generation in ultra-thin plasmonic metasurfaces without a full-metal backplane," Appl. Phys. B **124**, 1321–1327 (2018).

64. L. Tsang, D. Ahn, and S. L. Chuang, "Electric field control of optical second-harmonic generation in a quantum well," Appl. Phys. Lett. **52**, 697–699 (1988).

65. M. M. Fejer, S. J. B. Yoo, R. L. Byer, A. Harwit, and J. S. Harrisjr., "Observation of extremely large quadratic susceptibility at 9.6 lectric-field-biased AlGaAs quantum wells," Phys. Rev. Lett. **62**, 1041–1044 (1989).

66. E. Rosencher, A. Fiore, B. Vinter, V. Berger, P. Bois, and J. Nagle, "Quantum Engineering of Optical Nonlinearities," Science **271**, 168–173 (1996).

67. R. Sarma, J. X. U. Iaming, D. O. D. E. Ceglia, L. Uca, J. Xu, D. de Ceglia, L. Carletti, S. Campione, J. Klem, M. B. Sinclair, M. A. Belkin, and I. Brener, "An All-Dielectric Polaritonic Metasurface with a Giant Nonlinear Optical Response," Nano Lett. **22**, 896–903 (2022).

68. A. I. Kuznetsov, A. E. Miroshnichenko, Y. H. Fu, J. Zhang, and B. Luk'yanchuk, "Magnetic light," Sci. Rep. **2**, 4921–4926 (2012).

69. S. Jahani and Z. Jacob, "All-dielectric metamaterials," Nat. Nanotechnol. **11**, 23–36 (2016).

70. R. Camacho-Morales, M. Rahmani, S. Kruk, L. Wang, L. Xu, D. A. Smirnova, A. S. Solntsev, A. Miroshnichenko, H. H. Tan, F. Karouta, S. Naureen, K. Vora, L. Carletti, C. De Angelis, C. Jagadish, Y. S. Kivshar, and D. N. Neshev, "Nonlinear Generation of Vector Beams from AlGaAs Nanoantennas," Nano Lett. **16**, 7191–7197 (2016).

71. I. Staude and J. Schilling, "Metamaterial-inspired silicon nanophotonics," Nat. Photonics **11**, 274–284 (2017).

72. A. B. Evlyukhin, S. M. Novikov, U. Zywietz, R. L. Eriksen, C. Reinhardt, S. I. Bozhevolnyi, and B. N. Chichkov, "Demonstration of magnetic dipole resonances of dielectric nanospheres in the visible region," Nano Lett. **12**, 3749–3755 (2012).

73. P. A. Jeong, M. D. Goldflam, S. Campione, J. L. Briscoe, P. P. Vabishchevich, J. Nogan, M. B. Sinclair, T. S. Luk, and I. Brener, "High Quality Factor Toroidal Resonances in Dielectric Metasurfaces," ACS Photonics **7**, 1699–1707 (2020).

74. A. E. Miroshnichenko, A. B. Evlyukhin, Y. F. Yu, R. M. Bakker, A. Chipouline, A. I. Kuznetsov, B. Luk'yanchuk, B. N. Chichkov, and Y. S. Kivshar, "Nonradiating anapole modes in dielectric nanoparticles," Nat. Commun. **6**, 8069 (2015).

75. S. Campione, S. Liu, L. I. Basilio, L. K. Warne, W. L. Langston, T. S. Luk, J. R. Wendt, J. L. Reno, G. A. Keeler, I. Brener, and M. B. Sinclair, "Broken Symmetry Dielectric Resonators for High Quality Factor Fano Metasurfaces," ACS Photonics **3**, 2362–2367 (2016).

76. M. V. Rybin, K. L. Koshelev, Z. F. Sadrieva, K. B. Samusev, A. A. Bogdanov, M. F. Limonov, and Y. S. Kivshar, "High- Q Supercavity Modes in Subwavelength Dielectric Resonators," Phys. Rev. Lett. **119**, 2439011–2439015 (2017).

77. K. Koshelev, S. I. Lepeshov, M. Liu, A. Bogdanov, and Y. Kivshar, "Asymmetric Metasurfaces with High-Q Resonances Governed by Bound States in the Continuum," Phys. Rev. Lett. **121**, 1939031–1939035 (2018).

78. H. Friedrich and D. Wintgen, "Interfering resonances and bound states in the continuum," Phys. Rev. A **32**, 3231–3242 (1985).

79. A. C. Overvig, S. C. Malek, M. J. Carter, S. Shrestha, and N. Yu, "Selection rules for quasibound states in the continuum," Phys. Rev. B **102**, 0354341–03543428 (2020).

80. S. I. Azzam and A. V. Kildishev, "Photonic Bound States in the Continuum: From Basics to Applications," Adv. Opt. Mater. **9**, 20014691–200146913 (2021).

81. L. Carletti, K. Koshelev, C. De Angelis, and Y. Kivshar, "Giant Nonlinear Response at the Nanoscale Driven by Bound States in the Continuum," Phys. Rev. Lett. **121**, 339031–339035 (2018).

82. C. W. Hsu, B. Zhen, A. D. Stone, J. D. Joannopoulos, and M. Soljačić, "Bound states in the continuum," Nat. Rev. Mater. **1**, 160481–1604813 (2016).

83. J. D. Jackson, *Classical Electrodynamics*, 3rd ed. (Wiley, 1999).

84. S. Bergfeld and W. Daum, "Second-Harmonic Generation in GaAs: Experiment versus Theoretical Predictions of χ(2)xyz," Phys. Rev. Lett. **90**, 0368011–0368014 (2003).

85. S. Gehrsitz, F. K. Reinhart, C. Gourgon, N. Herres, A. Vonlanthen, and H. Sigg, "The refractive index of AlxGa1-xAs below the band gap: Accurate determination and empirical modeling," J. Appl. Phys. **87**, 7825–7837 (2000).

86. M. Decker, I. Staude, M. Falkner, J. Dominguez, D. N. Neshev, I. Brener, T. Pertsch, and Y. S. Kivshar, "High-

Efficiency Dielectric Huygens' Surfaces," Adv. Opt. Mater. **3**, 813–820 (2015).

87. I. Staude, A. E. Miroshnichenko, M. Decker, N. T. Fofang, S. Liu, E. Gonzales, J. Dominguez, T. S. Luk, D. N. Neshev, I. Brener, and Y. Kivshar, "Tailoring directional scattering through magnetic and electric resonances in subwavelength silicon nanodisks," ACS Nano **7**, 7824–7832 (2013).

88. S. Liu, G. A. Keeler, J. L. Reno, M. B. Sinclair, and I. Brener, "III–V Semiconductor Nanoresonators—A New Strategy for Passive, Active, and Nonlinear All-Dielectric Metamaterials," Adv. Opt. Mater. **4**, 1457–1462 (2016).

89. J. Sipe, D. Moss, and H. van Driel, "Phenomenological theory of optical second- and third-harmonic generation from cubic centrosymmetric crystals," Phys. Rev. B **35**, 1129–1141 (1987).

90. C. Yamada and T. Kimura, "Rotational symmetry of the surface second-harmonic generation of zinc-blende-type crystals," Phys. Rev. B **49**, 14372–14381 (1994).

91. R. Sanatinia, S. Anand, and M. Swillo, "Experimental quantification of surface optical nonlinearity in GaP nanopillar waveguides," Opt. Express **23**, 756–764 (2015).

92. R. Sanatinia, M. Swillo, and S. Anand, "Surface second-harmonic generation from vertical GaP nanopillars," Nano Lett. **12**, 820–826 (2012).

93. S. R. Armstrong, M. E. Pemble, A. Stafford, and A. G. Taylor, "Optical second-harmonic generation from GaAs(100) surfaces: The influence of H2," J. Phys. Condens. Matter **3**, S363–S366 (1991).

94. L. L. Hale, H. Jung, S. D. Gennaro, J. Briscoe, C. T. Harris, T. S. Luk, S. J. Addamane, J. L. Reno, I. Brener, and O. Mitrofanov, "Terahertz Pulse Generation from GaAs Metasurfaces," ACS Photonics **9**, 1136–1142 (2022).

95. K. Frizyuk, "Second-harmonic generation in dielectric nanoparticles with different symmetries," J. Opt. Soc. Am. B **36**, F32–F37 (2019).

96. K. Frizyuk, I. Volkovskaya, D. Smirnova, A. Poddubny, and M. Petrov, "Second-harmonic generation in Mie-resonant dielectric nanoparticles made of noncentrosymmetric materials," Phys. Rev. B **99**, 0754251–07542515 (2019).

97. A. Hayat, P. Ginzburg, and M. Orenstein, "Observation of two-photon emission from semiconductors," Nat. Photonics **2**, 238–241 (2008).

98. K. Thyagarajan, S. Rivier, A. Lovera, and O. J. F. Martin, "Enhanced second-harmonic generation from double resonant plasmonic antennae," Opt. Express **20**, 12860–12865 (2012).

99. M. Celebrano, X. Wu, M. Baselli, S. Großmann, P. Biagioni, A. Locatelli, C. De Angelis, G. Cerullo, R. Osellame, B. Hecht, L. Duò, F. Ciccacci, and M. Finazzi, "Mode matching in multiresonant plasmonic nanoantennas for enhanced second harmonic generation," Nat. Nanotechnol. **10**, 412–417 (2015).

100. L. Xu, M. Rahmani, D. Smirnova, K. Zangeneh Kamali, G. Zhang, D. Neshev, and A. Miroshnichenko, "Highly-Efficient Longitudinal Second-Harmonic Generation from Doubly-Resonant AlGaAs Nanoantennas," Photonics **5**, 29 (2018).

101. S. G. Lee, C. S. Kee, and S. H. Kim, "Bound states in the continuum (BIC) accompanied by avoided crossings in leaky-mode photonic lattices," Nanophotonics **9**, 4373–4380 (2020).

102. H. K. Gandhi, A. Laha, and S. Ghosh, "Ultrasensitive light confinement: Driven by multiple bound states in the continuum," Phys. Rev. A **102**, 0335281–0335285 (2020).

103. K. I. Okhlopkov, A. Zilli, A. Tognazzi, D. Rocco, L. Fagiani, E. Mafakheri, M. Bollani, M. Finazzi, M. Celebrano, M. R. Shcherbakov, C. De Angelis, and A. A. Fedyanin, "Tailoring Third-Harmonic Diffraction Efficiency by Hybrid Modes in High-Q Metasurfaces," Nano Lett. **21**, 10438–10445 (2021).

104. S. D. Gennaro, Y. Sonnefraud, N. Verellen, P. Van Dorpe, V. V. Moshchalkov, S. A. Maier, and R. F. Oulton, "Spectral interferometric microscopy reveals absorption by individual optical nanoantennas from extinction phase," Nat. Commun. **5**, 3748 (2014).

105. A. E. Miroshnichenko, S. Flach, and Y. S. Kivshar, "Fano resonances in nanoscale structures," Rev. Mod. Phys. **82**, 2257–2298 (2010).

106. M. Rahmani, B. Lukiyanchuk, B. Ng, A. Tavakkoli K. G., Y. F. Liew, and M. H. Hong, "Generation of pronounced Fano resonances and tuning of subwavelength spatial light distribution in plasmonic pentamers," Opt. Express **19**, 4949–4956 (2011).

107. K. Thyagarajan, J. Butet, and O. J. F. Martin, "Augmenting second harmonic generation using Fano resonances in plasmonic systems.," Nano Lett. **13**, 1847–1851 (2013).

108. N. Verellen, Y. Sonnefraud, H. Sobhani, F. Hao, V. V. Moshchalkov, P. Van Dorpe, P. Nordlander, and S. A. Maier, "Fano Resonances in Individual Coherent Plasmonic Nanocavities," Nano Lett. **9**, 1663–1667 (2009).

109. N. Karl, P. P. Vabishchevich, M. R. Shcherbakov, S. Liu, M. B. Sinclair, G. Shvets, and I. Brener, "Frequency Conversion in a Time-Variant Dielectric Metasurface," Nano Lett. **20**, 7052–7058 (2020).

110. J. Sautter, I. Staude, M. Decker, E. Rusak, D. N. Neshev, I. Brener, and Y. S. Kivshar, "Active tuning of all-dielectric metasurfaces," ACS Nano **9**, 4308–4315 (2015).

111. M. R. Shcherbakov, S. Liu, V. V. Zubyuk, A. Vaskin, P. P. Vabishchevich, G. Keeler, T. Pertsch, T. V. Dolgova, I. Staude, I. Brener, and A. A. Fedyanin, "Ultrafast all-optical tuning of direct-gap semiconductor metasurfaces," Nat. Commun. **8**, 17 (2017).

112. V. R. Almeida, C. A. Barrios, R. R. Panepucci, and M. Lipson, "All-optical control of light on a silicon chip," Nature **431**, 1081–1084 (2004).

113. K. Koshelev and Y. Kivshar, "Dielectric Resonant Metaphotonics," ACS Photonics **8**, 102–112 (2021).

114. D. Smirnova, D. Leykam, Y. Chong, and Y. Kivshar, "Nonlinear topological photonics," Appl. Phys. Rev. **7**, 0213061–02130626 (2020).

115. X. C. Zhang, Y. Jin, and X. F. Ma, "Coherent measurement of THz optical rectification from electro-optic crystals,"

Appl. Phys. Lett. **61**, 2764–2766 (1992).

116. F. Nastos and J. E. Sipe, "Optical rectification and shift currents in GaAs and GaP response: Below and above the band gap," Phys. Rev. B - Condens. Matter Mater. Phys. **74**, 1–15 (2006).

117. D. Coté, N. Laman, and H. M. Van Driel, "Rectification and shift currents in GaAs," Appl. Phys. Lett. **80**, 905–907 (2002).

118. U. A. Leon, D. Rocco, L. Carletti, M. Peccianti, S. Maci, G. Della Valle, and C. De Angelis, "THz-photonics transceivers by all-dielectric phonon-polariton nonlinear nanoantennas," Sci. Rep. **12**, 1–11 (2022).

119. O. Mitrofanov, T. Siday, R. J. Thompson, T. S. Luk, I. Brener, and J. L. Reno, "Efficient photoconductive terahertz detector with all-dielectric optical metasurface," APL Photonics **3**, (2018).

120. T. Siday, P. P. Vabishchevich, L. Hale, C. T. Harris, T. S. Luk, J. L. Reno, I. Brener, and O. Mitrofanov, "Terahertz Detection with Perfectly-Absorbing Photoconductive Metasurface," Nano Lett. **19**, 2888–2896 (2019).

121. L. L. Hale, C. T. Harris, T. S. Luk, S. J. Addamane, J. L. Reno, I. Brener, and O. Mitrofanov, "Highly efficient terahertz photoconductive metasurface detectors operating at microwatt-level gate powers," Opt. Lett. **46**, 3159 (2021).

122. J. Khurgin, "Second-order intersubband nonlinear-optical susceptibilities of asymmetric quantum-well structures," J. Opt. Soc. Am. B **6**, 1673–1682 (1989).

123. G. I. Stegeman, "χ(2) Cascading: nonlinear phase shifts," Quantum Semiclassical Opt. J. Eur. Opt. Soc. Part B **9**, 139–153 (1997).

124. G. I. Stegeman, D. J. Hagan, and L. Torner, "χ(2) Cascading Phenomena and Their Applications To All-Optical Signal Processing, Mode-Locking, Pulse Compression and Solitons.," Opt. Quantum Electron. **28**, 1691–1740 (1996).

125. P. Laffaille, J.-M. Manceau, T. Laurent, A. Bousseksou, L. Le Gratiet, R. Teissier, A. N. Baranov, and R. Colombelli, "Intersubband polaritons at λ ∼ 2 μ m in the InAs/AlSb system," Appl. Phys. Lett. **112**, 2011131–2011134 (2018).

126. A. S. Solntsev, G. S. Agarwal, and Y. Y. Kivshar, "Metasurfaces for quantum photonics," Nat. Photonics **15**, 327–336 (2021).

127. T. Santiago-Cruz, V. Sultanov, H. Zhang, L. A. Krivitsky, and M. V. Chekhova, "Entangled photons from subwavelength nonlinear films," Opt. Lett. **46**, 653–656 (2021).

128. G. Marino, A. S. Solntsev, L. Xu, V. F. Gili, L. Carletti, A. N. Poddubny, M. Rahmani, D. A. Smirnova, H. Chen, A. Lemaître, G. Zhang, A. V. Zayats, C. De Angelis, G. Leo, A. A. Sukhorukov, and D. N. Neshev, "Spontaneous photon-pair generation from a dielectric nanoantenna," Optica **6**, 1416–1422 (2019).

129. T. Santiago-Cruz, A. Fedotova, V. Sultanov, M. A. Weissflog, D. Arslan, M. Younesi, T. Pertsch, I. Staude, F. Setzpfandt, and M. Chekhova, "Photon Pairs from Resonant Metasurfaces," Nano Lett. **21**, 4423–4429 (2021).

130. M. Parry, A. Mazzanti, A. Poddubny, G. Della Valle, D. N. Neshev, and A. A. Sukhorukov, "Enhanced generation of nondegenerate photon pairs in nonlinear metasurfaces," Adv. Photonics **3**, 0550011–0550016 (2021).